# Subthreshold phonon generation in an optomechanical system with an exceptional point

A. Mukhamedyanov,[1] A. A. Zyablovsky,[1,2] E. S. Andrianov[1,2]

[1]*Moscow Institute of Physics and Technology, 141700, 9 Institutskiy pereulok, Moscow, Russia*
[2]*Dukhov Research Institute of Automatics (VNIIA), 127055, 22 Sushchevskaya, Moscow, Russia*

A phonon laser based on an optomechanical system consisting of two optical modes interacting with each other via a phononic mode is considered. An external wave exciting one of the optical modes plays a role of the pumping. It is shown that at some amplitude of the external wave an exceptional point exists. When the external wave amplitude is less than one corresponding to the exceptional point, the splitting of the eigenfrequencies takes place. It is demonstrated that in this case, the periodic modulation of the external wave amplitude can result in simultaneous generation of photons and phonons even below the threshold of optomechanical instability.

**Introduction.**

Optomechanical systems that combine optical and mechanical modes are the object of comprehensive study [1-4]. The interaction of light with mechanical modes makes it possible to implement new types of devices, e.g., phonon lasers [5-13]. Recently, it has been demonstrated that exceptional points (EPs) can be realized in the optomechanical systems [7-9,14,15]. The EP is a singularity in the eigenstate space of the system [16-18]. At parameters corresponding to the EP, two or more system eigenstates become linearly dependent and their eigenfrequencies coalesce [16,17]. The passing through the EP often leads to qualitative change in the system behavior, which is associated with a non-Hermitian phase transition [19-26]. For example, in parity-time (PT) symmetrical systems, the passing through the EP is accompanied by the spontaneous symmetry breaking in the eigenstates [19,20,27-30].

Systems with EPs are widely employed in applications. Unique properties of these systems caused by the EP are used to enhance sensitivity of the sensors [15,31-36] and laser gyroscopes [37-41], to select modes in the multimode lasers [28,42,43], and to achieve lasing without inversion [44,45]. There are many physical realizations of systems with EPs [19,20,27,44,46-49].

In this letter, we consider an optomechanical system of two optical modes interacting with each other via a phonon mode. The optical mode with greater frequency is excited by the external electromagnetic wave. There is a threshold value of the external wave amplitude, above which the simultaneous generation of photons and phonon occurs (laser/phonon laser). We determine the conditions at which the EP exists in such optomechanical system. We propose an approach to achieve the coherent generation below threshold of optomechanical instability. This method is based on the periodic modulation of the external wave amplitude near the EP. We suggest a way to create optical and phonon transistors based on the optomechanical systems with the EP.

**Model.** We consider an optomechanical system consisting of two modes of electromagnetic (EM) field with frequencies $\omega_1$ and $\omega_2$, respectively. These modes interact with each other via phonons with the frequency $\omega_b$. To describe this system, we use the following optomechanical Hamiltonian [5]

$$\hat{H} = \hbar\omega_1 \hat{a}_1^\dagger \hat{a}_1 + \hbar\omega_2 \hat{a}_2^\dagger \hat{a}_2 + \hbar\omega_b \hat{b}^\dagger \hat{b} + \hbar\Omega(\hat{a}_1^\dagger \hat{a}_2 \hat{b} + \hat{a}_1 \hat{a}_2^\dagger \hat{b}^\dagger) + \hbar\tilde{\Omega}(\hat{a}_1 e^{i\omega t} + \hat{a}_1^\dagger e^{-i\omega t}) \quad (1)$$

where $\hat{a}_{1,2}$ and $\hat{a}_{1,2}^\dagger$ are the annihilation and creation bosonic operators for the first and the second optical modes, respectively. $\hat{b}$ and $\hat{b}^\dagger$ are the annihilation and creation operators of the phonons that satisfy the bosonic communication relation $\left[\hat{b}, \hat{b}^\dagger\right] = 1$. The fourth term in (1) describes the optomechanical interaction between the electromagnetic modes via the phonons, $\Omega$ is a coupling strength between the modes and the phonons. The last term in the Hamiltonian describes the interaction of the first mode with the external EM wave. The intensity of the external wave is determined by $\tilde{\Omega}$.

To describe relaxation processes arising due to an interaction of the system with its environment, we use the master equation for density matrix $\hat{\rho}$ in the Lindblad form [50,51]. This equation includes Lindblad's superoperators describing the relaxation of the EM field modes and the phonons [52,53].

Using expressions $\langle \hat{A} \rangle = Tr(\hat{\rho}\hat{A})$ and $\left\langle \frac{d\hat{A}}{dt} \right\rangle = Tr\left(\frac{\partial \hat{\rho}}{\partial t}\hat{A}\right)$ we obtain equations for the average values of the operators $a_1 = \langle \hat{a}_1 \rangle$, $a_2 = \langle \hat{a}_2 \rangle$ and $b = \langle \hat{b} \rangle$:

$$\frac{da_1}{dt} = -(i\omega_1 + \gamma_1)a_1 - i\Omega a_2 b - i\tilde{\Omega}e^{-i\omega t} \quad (2)$$

$$\frac{da_2}{dt} = -(i\omega_2 + \gamma_2)a_2 - i\Omega a_1 b^* \quad (3)$$

$$\frac{db}{dt} = -(i\omega_0 + \gamma_0)b - i\Omega a_1 a_2^* \quad (4)$$

To obtain the closed system of differential equations we use the mean-field approximation [54] by making substitutions $\langle \hat{a}_1 \hat{b}^\dagger \rangle \to \langle \hat{a}_1 \rangle \langle \hat{b} \rangle^*$, $\langle \hat{a}_2 \hat{b} \rangle \to \langle \hat{a}_2 \rangle \langle \hat{b} \rangle$, $\langle \hat{a}_1 \hat{a}_2^\dagger \rangle \to \langle \hat{a}_1 \rangle \langle \hat{a}_2^\dagger \rangle$.

The Equations (2) - (4) describe the dynamics of the EM field and the phonons in the optomechanical system. To find the stationary solutions of the Equations (2)-(4) we make the following change of the variables $a_1 \to a_1 e^{-i\omega t}$, $a_2 \to a_2 e^{-i\omega t}$ and introduce the designation $\Delta_{1,2} = \omega_{1,2} - \omega$.

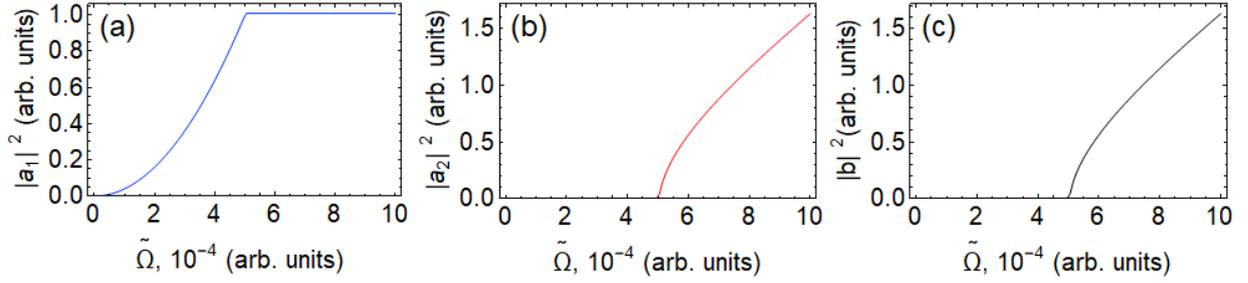

Figure 1. The dependence of $|a_1|^2$ (a), $|a_2|^2$ (b), $|b|^2$ (c) on the amplitude of the external EM wave ($\tilde{\Omega}$) exciting the first mode. Here $\gamma_1 = 10^{-2}\omega_1$, $\gamma_2 = \gamma_0 = 10^{-3}\omega_1$, $\Omega = 2\times 10^{-1}\omega_1$, $\omega = \omega_1$, $\omega_0 = 10^{-2}\omega_1$, $\omega_2 = \omega_1 + \omega_0$.

We demonstrate that the intensity of the EM field in the second mode $|a_2|^2$ and the intensity of phonons $|b|^2$ demonstrate the threshold dependence on the amplitude of the external EM wave exciting the first mode (i.e. $\tilde{\Omega}$) (Figure 1). In the case when the relaxation rates of $b$ and $a_2$ are equal ($\gamma_0 = \gamma_2$), the threshold value of the amplitude of the external wave $\tilde{\Omega}$ (a threshold of optomechanical instability) is determined by the following expression

$$\left|\tilde{\Omega}_{th}\right| = \frac{\sqrt{\Delta_1^2 + \gamma_1^2}}{|\Omega|} \frac{\sqrt{(\Delta_2 + \omega_0)^2 + (\gamma_0 + \gamma_2)^2}}{2} \qquad (5)$$

**Equations for amplitudes of small deviations from the stationary state**

Below the threshold of optomechanical instability (Figure 1), the stationary solution of the system is $a_1 = -\dfrac{\tilde{\Omega}}{(\omega_1 - \omega) - i\gamma_1}$, $a_2 = 0$ and $b = 0$. We study the stability of this solution. For this purpose, we derive the system of linear differential equations for amplitudes of small deviations from the stationary state

$$\frac{d(\delta a_2)^*}{dt} = (i\Delta_2 - \gamma_2)\delta a_2^* - i\frac{\Omega\tilde{\Omega}^*}{\Delta_1 + i\gamma_1}\delta b \tag{6}$$

$$\frac{d(\delta b)}{dt} = -(i\omega_0 + \gamma_0)\delta b + i\frac{\Omega\tilde{\Omega}}{\Delta_1 - i\gamma_1}\delta a_2^* \tag{7}$$

The eigenvalues of the Equations (6), (7) are given by the following expressions

$$\lambda_{1,2} = \frac{(i\Delta_2 - \gamma_2) - (i\omega_0 + \gamma_0)}{2} \pm \frac{1}{2}\sqrt{[(i\Delta_2 - \gamma_2) + (i\omega_0 + \gamma_0)]^2 + \frac{4|\Omega\tilde{\Omega}|^2}{\Delta_1^2 + \gamma_1^2}} \tag{8}$$

The zero solution of the Equation (2)-(4) is stable when the real parts of the eigenvalues are less than zero. The dependence of the eigenvalues on the amplitude of the external wave $\tilde{\Omega}$ is shown in Figure 2. We can see that in the system there is an exceptional point (EP), at which the eigenvalues are equal to each other and the corresponding eigenvectors coalesce. We obtain that the EP exists when $\gamma_0 = \gamma_2$ and $|\tilde{\Omega}| = |\tilde{\Omega}_{EP}| = \frac{\sqrt{\Delta_1^2 + \gamma_1^2}}{|\Omega|}\frac{\Delta_2 + \omega_0}{2}$. When $|\tilde{\Omega}| < |\tilde{\Omega}_{EP}|$, the imaginary parts of the eigenvalues are different and the system is in a strong coupling regime [20]. When $|\tilde{\Omega}| > |\tilde{\Omega}_{EP}|$, the system is in a weak coupling regime [20].

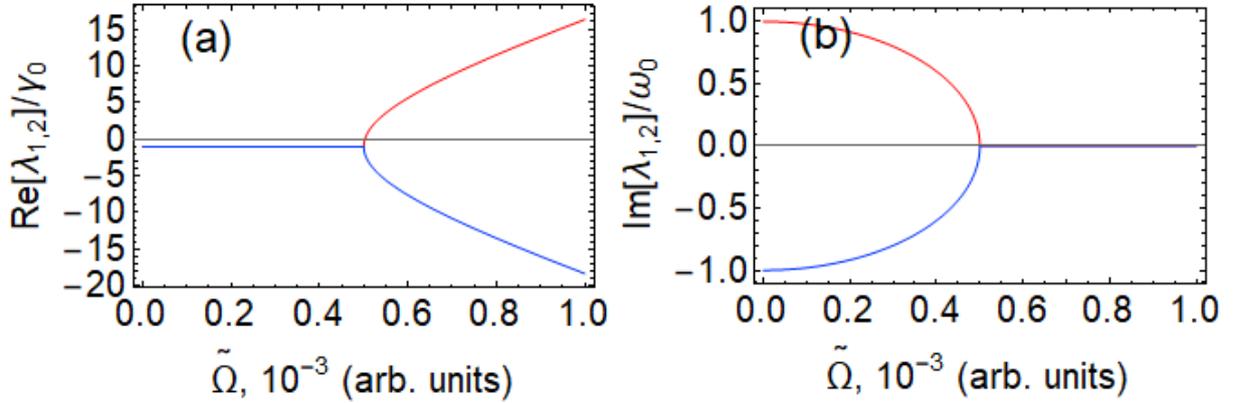

Figure 2. The dependence of the real (a) and imaginary (b) parts of the eigenvalues of the Equations (6), (7) on the amplitude of the external EM wave $\tilde{\Omega}$ exciting the first mode. Here $\gamma_1 = 10^{-2}\omega_1$, $\gamma_2 = \gamma_0 = 10^{-3}\omega_1$, $\Omega = 2\times 10^{-1}\omega_1$, $\Delta_1 = 0$, $\Delta_2 = 10^{-2}\omega_1$.

In the case, when the imaginary parts of the eigenvalues are not equal to each other the Equations (6), (7) can be reduced to the parametric oscillator equation [55]. To demonstrate it, we make the following substitutions and transformations $\delta a_2^* = A\exp\left(-i\left\{\frac{\omega_0-\Delta_2}{2}+\frac{\gamma_2+\gamma_0}{2}\right\}t\right)$, $\delta b = B\exp\left(-i\left\{\frac{\omega_0-\Delta_2}{2}+\frac{\gamma_2+\gamma_0}{2}\right\}t\right)$. As a result, we obtain the following equation

$$\ddot{A} - \left(\left(\frac{\gamma_0-\gamma_2}{2}+i\frac{\omega_0+\Delta_2}{2}\right)^2 + \frac{|\Omega\tilde{\Omega}|^2}{\Delta_1^2+\gamma_1^2}\right)A = 0 \qquad (9)$$

When $\gamma_0 = \gamma_2$ and $\omega_p^2 = \left(\left(\frac{\omega_0+\Delta_2}{2}\right)^2 - \frac{|\Omega\tilde{\Omega}|^2}{\Delta_1^2+\gamma_1^2}\right) > 0$ the Equation (9) coincides with the equation for the parametric oscillator, in which $\omega_p$ is played a role of the parametric oscillator frequency. Note that the condition $\omega_p^2 > 0$ ($\gamma_0 = \gamma_2$) is equivalent to the condition $|\tilde{\Omega}| < |\tilde{\Omega}_{EP}|$, that is, the system must be in the strong coupling regime.

In accordance with the theory of the parametric oscillator [55], the periodic time modulation of the frequency of the parametric oscillator allows for achieving the parametric amplification of the oscillations. Note that the Equations (6), (7) are reduced to the parametric oscillator equation when the amplitude of the external wave is less than the threshold value ($\tilde{\Omega} < \tilde{\Omega}_{th}$). This fact indicates that the oscillations in the considered optomechanical system can be induced below the threshold of optomechanical instability.

We use a periodic modulation of the amplitude of the external wave ($\tilde{\Omega}$) to excite the oscillations of the EM field in the second mode and the phonons below the threshold of optomechanical instability. We consider that the amplitude of the external EM wave exciting the first mode depends on time as

$$\tilde{\Omega} = \tilde{\Omega}_0\left(1+\alpha\sin(\Delta\omega t)\right) \qquad (10)$$

where $\tilde{\Omega}_0 < \tilde{\Omega}_{th}$.

The numerical simulation of the non-linear Equations (2)-(4) shows that such a modulation can excite the EM field in the second mode and the phonons (Figure 3). It occurs even when the average amplitude of the external wave is less than the threshold value.

The dependence $|a_2|^2$ and $|b|^2$ on the frequency $\Delta\omega$ and the amplitude $\alpha$ of the time modulation of $\tilde{\Omega}$ demonstrates the behavior that is corresponding to one in the parametric oscillator (Figure 4). It confirms the parametric nature of the excited oscillations in the considered optomechanical system.

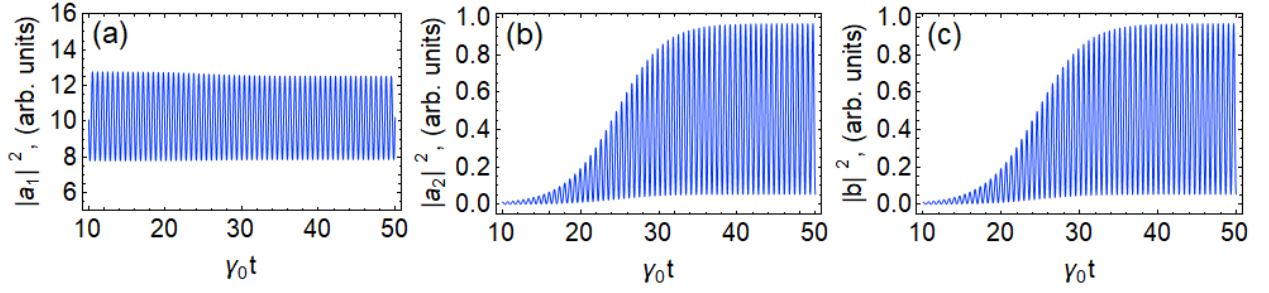

Figure 3. The temporal dependence of $|a_1|^2$ (a), $|a_2|^2$ (b), $|b|^2$ (c). The amplitude of the external EM wave $\tilde{\Omega}$ exciting the first mode periodically varies over time as $\tilde{\Omega} = \tilde{\Omega}_0 (1 + \alpha \sin(\Delta \omega t))$, where $\alpha = 0.17$, $\Delta \omega = 1.09 \omega_p$ and $\tilde{\Omega}_0 = 0.9 \tilde{\Omega}_{th}$ is less than the value of the laser threshold.

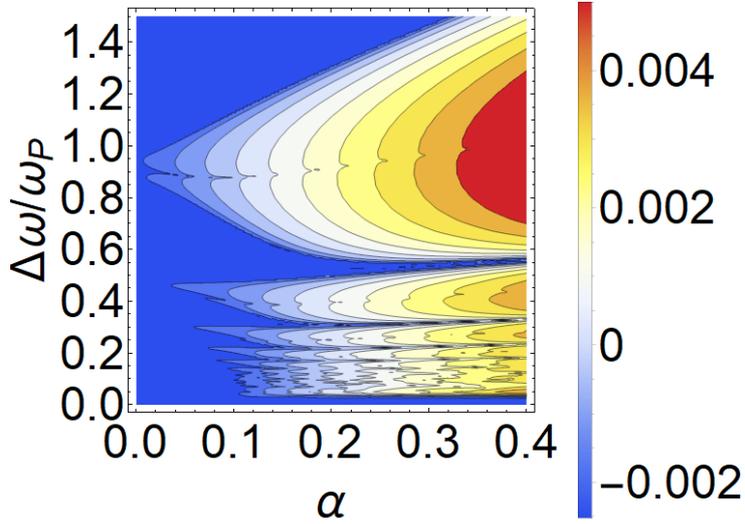

Figure 4. The dependence of the logarithm of the decrement of $|a_2|^2$ on the frequency $\Delta \omega$ and the amplitude of the modulation $\alpha$, when the amplitude of the external wave exciting the first mode, periodically varies over time as $\tilde{\Omega} = \tilde{\Omega}_0 (1 + \alpha \sin(\Delta \omega t))$. Here $\tilde{\Omega}_0 = 0.9 \tilde{\Omega}_{th}$, $\gamma_1 = 10^{-2} \omega_1$, $\gamma_2 = \gamma_0 = 10^{-3} \omega_1$, $\Omega = 2 \times 10^{-1} \omega_1$, $\Delta_1 = 0$, $\Delta_2 = 10^{-2} \omega_1$.

Thus, we conclude that the periodic modulation of the external wave amplitude can excite the EM field in the second mode and the phonons. Such a modulation can be used to create the lasers and the phonon lasers based on the optomechanical systems, in which the strong coupling regime takes place. The periodic modulation of the external wave amplitude that is necessary to operation of such lasers can be obtained by the interference of three electromagnetic waves, the frequency's difference of which is equal to $\Delta \omega$ (see Equation (10)). Indeed, if time evolutions of three waves are given as $E_1(t) = \tilde{\Omega}_0 \exp(-i\omega t)$, $E_2(t) = i\tilde{\Omega}_0 \frac{\alpha}{2} \exp(-i(\omega - \Delta\omega)t)$ and $E_3(t) = -i\tilde{\Omega}_0 \frac{\alpha}{2} \exp(-i(\omega + \Delta\omega)t)$, respectively. Then the time dependence for the external wave has form

$$E = E_1 + E_2 + E_3 = \tilde{\Omega}_0 \left(1 + \alpha \cos(\Delta\omega t)\right) \quad (11)$$

Thus, we obtain the necessary time dependence for the external wave. Note that the intensities of the second and third waves $\left|\tilde{\Omega}_0 \frac{\alpha}{2}\right|^2$ are much smaller than the intensity of the first wave $\left|\tilde{\Omega}_0\right|^2$ ($\alpha \ll 1$). At the same time, the parametric generation cannot occur without the second and third waves. Thus, it is possible to control the laser operation using low-intensity waves. This fact opens the way to the creation of optical and phonon transistors based on the considered optomechanical system with the EP.

**Conclusion.** We consider an optomechanical system consisting of two optical modes interacting with each other via the phonon mode. The external electromagnetic wave exciting the optical mode with greater frequency serves as a laser pumping. There is a threshold value of the external wave amplitude, above which the simultaneous generation of the optical and phonon modes occurs. We demonstrate that an exceptional point exists at a certain value of the external wave amplitude, which is less than the threshold of optomechanical instability. The exceptional point separates weak and strong coupling regimes. We propose the approach to achieve the generation of light and phonons in the subthreshold regime. We show the periodic modulation of the external wave amplitude can lead to excite the EM field and the phonons even below the threshold. A method for creating the optical and phonon transistors based on such an optomechanical system with the exceptional point is suggested.

Thus, our results open the way to the creation of lasers and phonon lasers generating in the subthreshold regime and the optical/phonon transistors on based them.

**Acknowledgments**

The study was financially supported by a Grant from Russian Science Foundation (project No. 22-72-00026).

**Funding**

The study was supported by a Grant from Russian Science Foundation (project no. 22-72-00026).

**References**

[1] T. J. Kippenberg and K. J. Vahala, Opt. Express **15**, 17172 (2007).
[2] T. J. Kippenberg and K. J. Vahala, Science **321**, 1172 (2008).
[3] M. Eichenfield, J. Chan, R. M. Camacho, K. J. Vahala, and O. Painter, Nature **462**, 78 (2009).
[4] M. Aspelmeyer, T. J. Kippenberg, and F. Marquardt, Rev. Mod. Phys. **86**, 1391 (2014).
[5] I. S. Grudinin, H. Lee, O. Painter, and K. J. Vahala, Phys. Rev. Lett. **104**, 083901 (2010).
[6] K. Vahala, M. Herrmann, S. Knünz, V. Batteiger, G. Saathoff, T. W. Hänsch, and T. Udem, Nature Phys. **5**, 682 (2009).
[7] H. Jing, S. K. Özdemir, X. Y. Lü, J. Zhang, L. Yang, and F. Nori, Phys. Rev. Lett. **113**, 053604 (2014).
[8] J. Zhang *et al.*, Nature Photon. **12**, 479 (2018).
[9] H. Lü, S. K. Özdemir, L. M. Kuang, F. Nori, and H. Jing, Phys. Rev. Appl. **8**, 044020 (2017).
[10] Y. Jiang, S. Maayani, T. Carmon, F. Nori, and H. Jing, Phys. Rev. Appl. **10**, 064037 (2018).
[11] R. M. Pettit, W. Ge, P. Kumar, D. R. Luntz-Martin, J. T. Schultz, L. P. Neukirch, M. Bhattacharya, and A. N. Vamivakas, Nature Photon. **13**, 402 (2019).
[12] D. L. Chafatinos, A. S. Kuznetsov, S. Anguiano, A. E. Bruchhausen, A. A. Reynoso, K. Biermann, P. V. Santos, and A. Fainstein, Nature Commun. **11**, 4552 (2020).
[13] J. Kabuss, A. Carmele, T. Brandes, and A. Knorr, Phys. Rev. Lett. **109**, 054301 (2012).
[14] H. Xu, D. Mason, L. Jiang, and J. G. E. Harris, Nature **537**, 80 (2016).
[15] P. Djorwe, Y. Pennec, and B. Djafari-Rouhani, Phys. Rev. Appl. **12**, 024002 (2019).
[16] M. V. Berry, Czech. J. Phys. **54**, 1039 (2004).
[17] N. Moiseyev, *Non-Hermitian Quantum Mechanics* (Cambridge University Press, Cambridge, UK, 2011).
[18] W. D. Heiss, J. Phys. A **45**, 444016 (2012).
[19] M.-A. Miri and A. Alù, Science **363**, eaar7709 (2019).
[20] S. Özdemir, S. Rotter, F. Nori, and L. Yang, Nat. Mater. **18**, 783 (2019).
[21] C. M. Bender and S. Boettcher, Phys. Rev. Lett. **80**, 5243 (1998).
[22] C. M. Bender, S. Boettcher, and P. N. Meisinger, J. Math. Phys. **40**, 2201 (1999).
[23] S. Klaiman, U. Günther, and N. Moiseyev, Phys. Rev. Lett. **101**, 080402 (2008).
[24] A. V. Sadovnikov, A. A. Zyablovsky, A. V. Dorofeenko, and S. A. Nikitov, Phys. Rev. Appl. **18**, 024073 (2022).
[25] T. T. Sergeev, A. A. Zyablovsky, E. S. Andrianov, and Y. E. Lozovik, arXiv **2207.01862** (2022).
[26] T. T. Sergeev, A. A. Zyablovsky, E. S. Andrianov, A. A. Pukhov, Y. E. Lozovik, and A. P. Vinogradov, Sci. Rep. **11**, 24054 (2021).
[27] R. El-Ganainy, K. G. Makris, M. Khajavikhan, Z. H. Musslimani, S. Rotter, and D. N. Christodoulides, Nat. Phys. **14**, 11 (2018).
[28] L. Feng, Z. J. Wong, R.-M. Ma, Y. Wang, and X. Zhang, Science **346**, 972 (2014).
[29] K. G. Makris, R. El-Ganainy, D. N. Christodoulides, and Z. H. Musslimani, Phys. Rev. Lett. **100**, 103904 (2008).
[30] C. E. Rüter, K. G. Makris, R. El-Ganainy, D. N. Christodoulides, M. Segev, and D. Kip, Nature Phys. **6**, 192 (2010).

[31] J. Wiersig, Phys. Rev. A **93**, 033809 (2016).
[32] J. Wiersig, Photonics Res. **8**, 1457 (2020).
[33] W. Chen, Ş. Kaya Özdemir, G. Zhao, J. Wiersig, and L. Yang, Nature **548**, 192 (2017).
[34] M. Zhang, W. Sweeney, C. W. Hsu, L. Yang, A. D. Stone, and L. Jiang, Phys. Rev. Lett. **123**, 180501 (2019).
[35] W. Langbein, Phys. Rev. A **98**, 023805 (2018).
[36] H. Hodaei, A. U. Hassan, S. Wittek, H. Garcia-Gracia, R. El-Ganainy, D. N. Christodoulides, and M. Khajavikhan, Nature **548**, 187 (2017).
[37] Y.-H. Lai, Y.-K. Lu, M.-G. Suh, Z. Yuan, and K. Vahala, Nature **576**, 65 (2019).
[38] M. P. Hokmabadi, A. Schumer, D. N. Christodoulides, and M. Khajavikhan, Nature **576**, 70 (2019).
[39] S. Sunada, Phys. Rev. A **96**, 033842 (2017).
[40] H. Wang, Y. H. Lai, Z. Yuan, M. G. Suh, and K. Vahala, Nature Commun. **11**, 1610 (2020).
[41] X. Mao, G. Q. Qin, H. Yang, H. Zhang, M. Wang, and G. L. Long, New J. Phys. **22**, 093009 (2020).
[42] H. Hodaei, M.-A. Miri, M. Heinrich, D. N. Christodoulides, and M. Khajavikan, Science **346**, 975 (2014).
[43] A. A. Zyablovsky, I. V. Doronin, E. S. Andrianov, A. A. Pukhov, Y. E. Lozovik, A. P. Vinogradov, and A. A. Lisyansky, Laser Photonics Rev. **15**, 2000450 (2021).
[44] I. V. Doronin, A. A. Zyablovsky, E. S. Andrianov, A. A. Pukhov, and A. P. Vinogradov, Phys. Rev. A **100**, 021801(R) (2019).
[45] I. V. Doronin, A. A. Zyablovsky, and E. S. Andrianov, Opt. Express **29**, 5624 (2021).
[46] S. Longhi, Europhys. Lett. **120**, 64001 (2018).
[47] A. A. Zyablovsky, E. S. Andrianov, and A. A. Pukhov, Sci. Rep. **6**, 29709 (2016).
[48] E. Peter, P. Senellart, D. Martrou, A. Lemaitre, J. Hours, J. M. Gerard, and J. Bloch, Phys. Rev. Lett. **95**, 067401 (2005).
[49] D. Zhang, X.-Q. Luo, Y.-P. Wang, T.-F. Li, and J. Q. You, Nat. Commun. **8**, 1368 (2017).
[50] C. Gardiner and P. Zoller, *Quantum noise: a handbook of Markovian and non-Markovian quantum stochastic methods with applications to quantum optics* (Springer Science & Business Media, 2004).
[51] H. Carmichael, *An open systems approach to quantum optics* (Springer-Verlag, Berlin, 1991).
[52] J. Kabuss, A. Carmele, and A. Knorr, Phys. Rev. B **88**, 064305 (2013).
[53] V. Y. Shishkov, E. S. Andrianov, A. A. Pukhov, A. P. Vinogradov, and A. A. Lisyansky, Phys. Rev. A **100**, 053838 (2019).
[54] M. O. Scully and M. S. Zubairy, *Quantum optics* (Cambridge University Press, 1999).
[55] L. D. Landau and E. M. Lifshitz, *Mechanics* (Butterworth-Heinemann, 1976), 3rd edn., Vol. 1, Course of Theoretical Physics,  p.^pp. 200.